# Specifying and Reasoning about Contextual Preferences in the Goal-oriented Requirements Modelling

Khavee Agustus Botangen[1], Jian Yu[1], Sira Yongchareon[1], Liang Huai Yang[2], Quan Bai[1]

[1]School of Engineering, Computer and Mathematical Sciences, Auckland University of Technology, New Zealand
{khavee.botangen, jian.yu sira.yongchareon, quan.bai}@aut.ac.nz
[2]School of Computer Science and Technology, Zhejiang University of Technology, China
yang.lianghuai@gmail.com

## ABSTRACT

Goal-oriented requirements variability modelling has established the understanding for adaptability in the early stage of software development – the Requirements Engineering phase. Goal-oriented requirements variability modelling considers both the intentions, which are captured as goals in goal models, and the preferences of different stakeholders as the main sources of system behaviour variability. Most often, however, intentions and preferences vary according to contexts. In this paper, we propose an approach for a contextual preference-based requirements variability analysis in the goal-oriented Requirements Engineering. We introduce a quantitative contextual preference specification to express the varying preferences imposed over requirements that are represented in the goal model. Such contextual preferences are used as criteria to evaluate alternative solutions that satisfy the requirements variability problem. We utilise a state-of-the-art reasoning implementation from the Answer Set Programming domain to automate the derivation and evaluation of solutions that fulfill the goals and satisfy the contextual preferences. Our approach will support systems analysts in their decisions upon alternative design solutions that define subsequent system implementations.

## CCS CONCEPTS

• **Software and its engineering** ~ **Requirements analysis** • Software and its engineering ~ Software design engineering

## KEYWORDS

Requirements engineering, contextual preferences, goal modelling, variability, adaptability, context-aware

## 1 INTRODUCTION

Requirements variability, which is typically characterised by creating multiple subsets of requirements, has long been advocated in Requirements Engineering (RE) to establish a *problem-space-oriented* approach to software adaptability [5], [17], [9]. Each subset, called a *requirements variant*, describes a candidate solution to the general requirements variability problem: deriving which requirements to achieve, as well as which alternatives to adopt in reaching those requirements.

*Preferences* have become an important factor to consider in this challenge of requirements variability analysis. Various preference-based goal models have been proposed to support the goal-oriented approach of requirements specification. Such models represent requirements as goals which are classified as either *hardgoals* or *softgoals*. In the context of this paper, hardgoals are the mandatory stakeholder requirements that need to be fulfilled. Each hardgoal has a *means* of fulfilling it, and such means is likewise considered a hardgoal. In a given goal model, a set of these means compose a requirements variant, defining a candidate solution to the depicted requirements problem. Softgoals, on the other hand, are the optional requirements often expressed as high-level quality requirements.

Traditionally, *qualitative* or *quantitative* preference valuations specified over the softgoals are used to distinguish the requirements variants. With the qualitative preference valuation, preferences between any of the softgoals are directly specified, typically as a binary relation. This presents a natural way of expressing the desirability of one softgoal over another. A qualitative approach can directly model an explicit expression of a stakeholder's preference, such as: "We prefer maintaining patient's independence than providing better comfort". In this regard, we take for example a goal such as *patient takes medication* as one of the hardgoals of a personal medication assistant application. There are two means of fulfilling this hardgoal: a *self-initiated* intake which can encourage a patient to be independent, or a *robot-assisted* intake which can provide a patient with better comfort. Consequently, solutions containing the former would be more desirable than those with the latter. With approaches using quantitative preferences, goal models integrate scoring functions associating numerical scores over softgoals to indicate prioritisation by stakeholders. Such numerical preferences are more precise and intuitively understandable as long as there is a readily available knowledge about the degree to which one softgoal is desired over another. For instance, the softgoals that the stakeholders may be interested to satisfy could be ranked by assigning scores based on importance.

Most often, preferences vary. Different stakeholders can have different preferences in different situations. We refer to such situations as *contexts*, which are facts that capture the nature of the system's operational environment. Referring again to the means of achieving the goal *patient takes medication*: a *robot-assisted* or a *self-initiated* (non-robot assisted) intake, varying preferences can be imposed between these two alternatives. For instance, when the patient has dementia, a robot-assisted intake may be preferred, whereas when the patient goes outdoors, a self-initiated intake is necessary. A patient may also state particular preferences such as, "*I prefer robot assistance when I am busy or when I'm tired*", "*I believe I'm still strong and capable of taking the medication on my own, but it would be nice to use the robot when my rheumatism attacks as it'd be difficult for me to move*", or "*As long as the medicine is within my reach, I might not need the robot*". It is worth noting that variations in stakeholder preferences depend on context. However, such contextual preference variability has been given little attention in the literature.

In this paper, we propose an approach for a contextual preference-based requirements variability analysis in the goal-oriented RE. We propose a *contextual preference model* that enhances preferences with contextual information to capture the changing nature of stakeholders' desires caused by context variation. Here, we capture two ways of expressing preferences. First, stakeholders may attribute different levels of importance to softgoals. Preferences are therefore imposed over such goals. Second, stakeholders may also express preferences over the alternatives, i.e., a means to achieving a goal is preferred over another alternative. Hence, preferences are not only asserted for softgoals but may also be expressed for hardgoals. As such, a stakeholder may convey their preferences explicitly, particularly among alternative hardgoals. We then propose a reasoning technique that applies the contextual preferences to derive and evaluate requirements variants. Our approach can guide analysts in their decisions about alternative design solutions that will best satisfy a requirements problem.

## 2 BACKGROUND AND MOTIVATION

Goal-Oriented Requirements Engineering (GORE) [26], [19] has become a mainstream technique in RE that has shifted the perspective of requirements from *functional* to *intentional*. The functional perspective is reflected in traditional structured analysis techniques and considers requirements as functions that the system should support. The intentional perspective views requirements as the main expression of stakeholders' intentions and explicitly represents the *why* (intended purpose) of the system [13], [8], [7]. In GORE, goal models capture and refine stakeholders' *goals*, where such goals represent the requirements of a system.

Our attention to requirements variability focuses on finding solutions to a given requirements problem. We consider a *solution* as a combination of tasks that must be operationalised to achieve the overall goal, i.e., the root goal. A considerable source of requirements variability are preferences. In RE practice, managing requirements has been strongly influenced by optionality and prioritisation [7]. In this regard, goal modelling classifies requirements into two types, based on their optionality among potential solutions: *mandatory* requirements are the necessary ones to fulfill the root goal, and *optional* requirements are desired but not necessary to fulfill the root goal [16], [7]. A mandatory requirement is represented by a hardgoal while an optional requirement is the nice-to-have one, represented by a softgoal. In order to become acceptable, a solution has to fulfill the mandatory requirements. An acceptable solution does not have to achieve an optional requirement, but if it does, such a solution becomes more desirable than those which do not. We aim to understand how contextual preferences posed by stakeholders over these requirements impact the variability of the alternative choices that will comprise a solution.

### 2.1 Overview of Goal Modelling

A typical goal model is an annotated AND/OR graph showing how higher-level goals are satisfied by lower-level ones (i.e., goal refinement), and conversely, how lower-level goals contribute to the fulfilment of higher-level ones (i.e., goal abstraction) [27]. We show in Figure 1 a partial goal model that specifies the requirements of a personal medication assistant further described in Section 2.2. The main elements of a goal model are *goals*. Goals, represented by the oval and cloud shapes, prescribe the intents (i.e., state of affairs or conditions) the system or actors of interest would like to satisfy. The leaf-level goals are called *tasks* – those in the hexagonal shapes – and define specific activities performed by the system or actors to operationalise and fulfill their goals. The relationships among goals and tasks are mainly depicted by *means-end links* ($\longrightarrow$). There is a means (i.e., sub-goal or task) to fulfill a certain end (i.e., parent goal). Multiple means are associated with the *AND-/OR-decomposition links* ($\xrightarrow{AND}$/$\xrightarrow{OR}$). An AND-decomposed goal implies that the satisfaction/performance of each of its children is necessary for it to be fulfilled. An OR-decomposed goal indicates that the satisfaction/performance of any of its children is sufficient to satisfy the parent goal.

Goals are mainly classified as either *hardgoals* or *softgoals*. A hardgoal, also called a behavioural goal [27], declaratively prescribes an intended system behaviour. It is denoted by an oval shape in Figure 1. A hardgoal has a clear-cut criterion to decide whether it is satisfied or not, e.g., a set of system operations can be performed to satisfy such goals. For instance, the goal *track medication history* ($g_2$) is satisfied by performing the tasks: *patient confirms intake*, *auto-monitoring of vital signs*, and *inform relatives*. In contrast, a softgoal (denoted by the cloud shape) cannot be established in a clear-cut sense. It can be satisfied to a 'good enough' degree, depending on subjective judgement and based on relevant evidence. For the personal medication assistant, we cannot say in a strict sense whether a certain system behaviour satisfies or not the goal of *minimal patient effort*. However, we may say that one system behaviour contributes further to minimising the patient's effort than another. Thus, the degree of satisfaction of a softgoal may be higher or lower when comparing between the alternatives. Because goal satisfaction should not be judged in a strict sense with softgoals, the phrase *goal satisficing* is sometimes used instead [27]. Relationships between hardgoals and softgoals are then established by the *contribution links* – the *make* ($\xrightarrow{++}$) and *break* ($\xrightarrow{--}$) links, denoted by the dashed arrow lines in the figure. The *make* link means that satisfying or performing the origin hardgoal or task satisfices the target softgoal. The *break* link means that satisfying or performing the origin hardgoal or task denies satisficing the target softgoal. Nonetheless in both cases, if the origin is neither satisfied nor performed, this does not imply anything about the target. Through such correlations, the different alternatives that contribute to different degrees of satisfying the softgoals can be reflected.

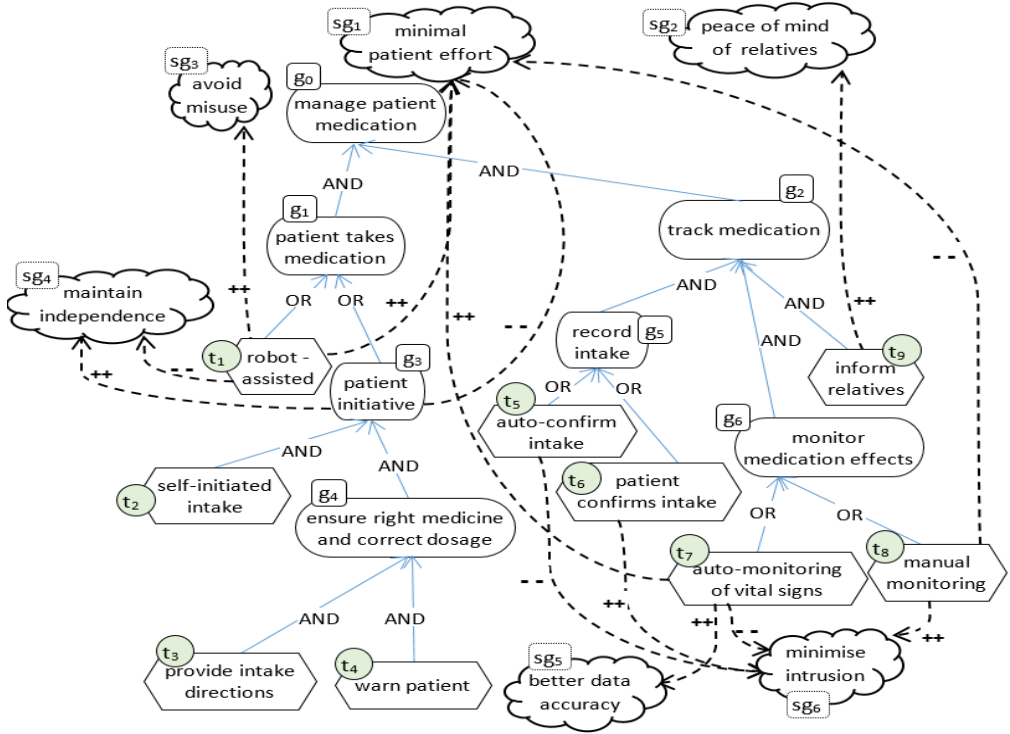

**Figure 1: A goal model for a personal medication assistant.**

A goal model can be represented formally so that each goal $g$ is associated with a propositional literal and the satisfaction of the root goal is represented by the propositional formula $G \equiv T_g \wedge Q_g$. This Propositional Calculus equivalent formalism is similarly used in [15] and can directly apply to the diagrammatic formalism in our goal model. The formula $T_g$ denotes the AND/OR structure in terms of leaf level literals, i.e., tasks. $Q_g$ is the additional constraint links, i.e., contribution links. Given two goals $g_1$ and $g_2$, the constraint links $g_1 \xrightarrow{++} g_2$ and $g_1 \xrightarrow{--} g_2$ respectively result as conjuncts $g_1 \rightarrow g_2$ and $g_1 \rightarrow \neg g_2$ in the formula $Q_g$. In both $T_g$ and $Q_g$, a literal representing a non-leaf node is recursively replaced by its AND/OR decomposition until a clause that contains only leaf level nodes is reached. For instance, for the goal $g_1$ in Figure 1: $T_{g1} \equiv t_1$ OR (($t_2$ AND ($t_3$ AND $t_4$))) and $Q_{g1} \equiv (t_1 \rightarrow sg_3)$ AND $(t_1 \rightarrow sg_1)$ AND $(t_1 \rightarrow \neg sg_4)$ AND (($t_2$ AND ($t_3$ AND $t_4$)) $\rightarrow sg_4$) AND $(t_1 \rightarrow \neg sg_4)$ AND (($t_2$ AND ($t_3$ AND $t_4$)) $\rightarrow \neg sg_1$). Hence, an alternative solution for a goal model would satisfy the resultant propositional formula $G$.

Our goal model adopts a subset of the i* goal modelling language originally proposed in [29]. Multiple variants of this language have been proposed and an effort towards its standardisation is already initiated, i.e., iStar 2.0 [4]. In our example in Figure 1, we omit some fundamental language constructs for simplicity. For example, some of the leaf-level tasks can be delegated (i.e., via *dependency links*) to other actors such as external services and systems, e.g., the task *inform relatives* may be delegated to a service provided by a mobile notification application. Moreover, we minimise refinements of the leaf-level goals (i.e., considered as tasks), even though they could be further AND-/OR-decomposed into lower level forms such as by applying the variability frames approach introduced in [15]. For instance, the task of *informing relatives* could be refined by considering the concern of the information, whether it's a *problem alert* or an *intake notification*. Likewise, the manner to *provide intake directions* (e.g., through *mobile phone* or *medicine cabinet screen*) or the different ways of *manual monitoring* of medication side effects (e.g., *patient manually*

*inputs observation*, *tele-consultation*, or *a visit to the physician*) could be represented.

## 2.2 Motivating Scenario: Personal Medication Assistant

We consider a requirements scenario from the RAMCIP (Robotic assistant for MCI patients) project [12], particularly the function of providing assistance for taking medications. We describe an assistive medication system, i.e., personal medication assistant, which facilitates patients' medication routine, both to ensure that a patient takes the needed medication, and to monitor proper medication or food supplement intake. We enumerate some objectives and functions of the system. It gives reminders when an intake is missed, or when a patient is leaving home and has to take some pills. It provides intake directions (e.g., correct dosage), or a warning at an attempt to take the wrong pill. It keeps a record of the patient's medicine intakes and monitors medication results and side-effects. To maintain peace of mind and the feeling of assurance, an intake is conveyed to relatives. Similarly, for problems such as consecutive misses or negative side-effects, the system informs external parties like relatives or caregivers. In order to perform such functions, the system utilises appropriate services from objects within the smart home of the patient. For example, a reminder uses the display service of the kitchen TV when the patient is in the kitchen. The capability of the patient should also be considered by the system. When a patient with a mild hearing impairment is in their backyard lawn, a reminder utilising an audio speaker service is adjusted to a higher volume to ensure being noticed. At times when a patient is tired or unable to walk to the medicine cabinet, the medicine dispenser is fetched by a robot service. But when an accompanying person (e.g., relative or caregiver) is present, the use of a robot should be restrained as human assistance is preferred for self-initiated intake. We illustrate in Figure 1 a partial requirements goal model of the personal medication assistant. A candidate solution, which satisfies the overall requirements problem posed by the root goal, is a set of tasks. For instance, the tasks [$t_1$, $t_5$, $t_7$, $t_9$] and [$t_3$, $t_4$, $t_2$, $t_5$, $t_8$, $t_9$] are each a candidate solution to the requirements problem defined by the root goal $g_0$. A solution strictly satisfies the AND/OR structure. In contrast, neither the tasks [$t_6$, $t_7$, $t_9$] nor [$t_2$, $t_3$, $t_4$, $t_6$, $t_7$] are acceptable solutions. Neither satisfies the AND/OR decomposition, i.e., the goal $g_1$ is not achieved in the former, and the latter failed to satisfy the AND refinement of $g_2$.

From the aforementioned requirements scenario, we observe the potential alternatives of behavioural designs that can be specified to meet every system objective. These alternatives (*aka. variabilities*) come as the OR-decompositions in the goal model, e.g., $t_1$ or $g_3$ are alternatives to fulfill $g_1$. We call an OR-decomposed goal as *variability point*, e.g., $g_1$. Because operationalising such large space of alternatives would be onerous, it becomes practical to identify the more suitable ones to comprise a solution from which conforming design decisions can be derived. The suitability of a solution may depend on either *context instances* or the *criteria given by stakeholders*, or both. On the one hand, the applicability of each alternative can be context dependent. On the other hand, various stakeholders, like patients, care givers, relatives, physicians, and organisations providing services or goods, may have different prioritisation over *i)* the elements comprising the solution set (alternative hardgoals) or *ii)* the high-level objectives of the alternative solutions (softgoals). Moreover, even the same stakeholder, e.g., an individual patient, can have different priorities in different times and situations. Overall, this requirements scenario raises the question, "Which among the behavioural design alternatives are suitable to the stakeholders' preferences in a given situation and context". Accordingly, there is a need to represent both context instances and stakeholders' criteria as means to explicitly characterise every solution set.

## 3 THE CONTEXTUAL PREFERENCE MODEL

In this section, we present a model that associates preferences with contexts. We enhance a quantitative preference specification over goals, e.g., [16], with contextual annotations. In particular, we build our model from the approach proposed by Stefanidis et al. in [23].

### 3.1 Context Specification

It is important to show how to specify the contexts in which preferences apply. Given the system's high-level goals and their refinements, we identify a finite set of entities comprising the operational environment.

**Definition 1** (Context dimension). Given a system *S*, the context dimension of *S*, $CD_S$, is a finite set of entities, $CD_S = \{C_1, C_2, ..., C_n\}$. Each entity $C_i$, $1 \le i \le n$, is called a context element.

Context dimension is a set of environment entities relevant to the system. Each entity represents a physical or conceptual object that can take observable values to describe a particular contextual instance. For example, take the personal medication assistant. We identify several relevant context elements: *patient_activity*, *patient_location*, *patient_illness*, *weather*, *body_condition*, and *accompanying_people*. Hence, its context dimension is $CD_{PMA}$ = {*patient_activity*, *patient_location*, *patient_illness*, *weather*, *body_condition*, *accompanying_people*}. We show in Table 1 context elements composing the context dimension of our personal medication assistant. Each element has an enumerated domain of values. Here, we adopt the representational view for context described in [6]. We note that context values are data the system has to capture from its environment. Methods of capturing context, however, are beyond the scope of this paper.

**Definition 2** (Context instance). Given a context dimension $CD_S = \{C_1, C_2, ..., C_n\}$, a context instance is an *n*-tuple of the form $(c_1, c_2, ..., c_n)$, $c_i \in dom(C_i)$, where $dom(C_i)$ is the domain of values for $C_i$, $1 \le i \le n$.

A context instance is the state of fact that describes the environment by assigning values to the elements of a context dimension. For our personal medication assistant, *(busy, outdoor, dementia, good, normal, alone)* and *(busy, near_dispenser,*

Table 1: Context elements and their values.

| Context element | Values |
| --- | --- |
| patient_activity | busy, idle |
| patient_location | indoor, outdoor, kitchen, living_room, near_dispenser |
| patient_illness | dementia, MCI, normal |
| weather | bad, good |
| body_condition | sick, tired, normal |
| accompanying_people | caregiver, relatives, alone |

*dementia, good, tired, caregiver)* are some of its context instances. The set of all possible context instances $W$ is the Cartesian product of the domains of the context elements, i.e., $W = dom(C_1) \times dom(C_2) \times ... \times dom(C_n)$.

**Definition 3** (Context assertion). Let $CD_S = \{C_1, C_2, ..., C_n\}$ be a context dimension with $C_i$, $1 \le i \le n$, as its elements. A context assertion $con(C_i)$ is an expression of the form $C_i \in \{c_1, ...., c_r\}$, where $c_q \in dom(C_i)$, $1 \le q \le r$.

A combined context assertions of the elements in a context dimension is used to specify context instances. This combination, which we call *con*, is an expression $(con(C_1) \land ... \land con(C_n))$, where $con(C_i)$ is a context assertion for a single element $C_i \in CD_S$ and there is at most one context assertion for $C_i$. If the context assertions of all context elements do not appear in *con*, the values of the missing element are considered indifferent. Hence, whenever $C_i$ is not found in *con*, an assertion $C_i \in \{All\}$ is implicitly included as part of *con*, where *All* is a value that denotes all the values from the domain $dom(C_i)$ of context element $C_i$. For example, given $CD_{PMA}$, the combined context assertion $con = (patient\_location \in \{indoor\} \land weather \in \{good\} \land body\_condition \in \{tired, sick\})$ specifies two context instances: *(All, indoor, All, good, tired, All)*, and *(All, indoor, All, good, sick, All)*. Essentially in this example, *con* would specify a set of 36 context instances defined by the Cartesian product $V_1 \times V_2 \times ... \times V_n$, where $V_i = \{c_1, ...., c_r\}$ for each context assertion $C_i \in \{c_1, ...., c_r\}$. However, the use of the value *All* reduces this number of context instances to only two.

### 3.2 Contextual Preference Specification

Preference specification has two general approaches: a qualitative and a quantitative preference. The qualitative approach directly specifies preferences between objects of concern, typically by using binary preference relations. The quantitative approach indirectly specifies preferences, using scoring functions that associate a numeric score to the objects of concern. Contextual preference specification, which explicitly indicates the context instance a preference holds, can apply to both approaches. In this work, we use a quantitative contextual preference model in which we annotate preferences with both contexts and scoring components.

**Definition 4** (Contextual preference). A contextual preference $p$ is a triple (*Action, con, score*).

- *Action* is a satisfaction expression of a desired property in the solutions implied by the goal model. *Action* is in the form *satisfy(g)* or *perform(t)* respectively denoting the preference to the satisfaction of goal $g$ or performance of task $t$.
- *con* is a (combined) context assertion.
- *score* is a numerical value in the range [0,n].

Our contextual preference specification employs a scoring mechanism that follows the quantitative preference framework of Agrawal and Wimmers [1], which was also used in [10] and [23]. The scoring component *score* indicates the degree of prioritisation assigned to an *Action* considering the set of context instances specified by *con*. The value $n$, which in this paper is set to 10, means extreme interest and the value 0 means no interest. The contextual preference specification allows the expression of priorities through priority rankings of the desired properties of expected solutions, as well as considering the circumstances that hold for such priorities. We take as an example the goal model of our personal medication assistant. The contextual preference *(perform(robot-assisted), patient_illness ∈ {dementia}, 9)* expresses that when the patient has *dementia*, performing the task *robot-assisted* for taking a medication is preferred with a priority score of 9.

Multiple satisfaction expressions that are given similar preference scores on a particular situation can be combined in a single contextual preference notation. The contextual preference $p_1$ in Figure 2 combines the three preferences for *perform($t_1$)*, *perform($t_5$)*, and *perform($t_7$)*, and each one is similarly preferred with a score of 9 when the patient's illness is dementia. The symbol • is used as a separator in this shortcut notation.

**Definition 5** (Preference catalogue). A *preference catalogue P* is a set of contextual preferences that applies to a particular requirements goal model.

A preference catalogue, through its set of contextual preferences, defines the priorities imposed over the hardgoal alternatives and softgoals in a certain situation or context. A catalogue may contain the desires of an individual stakeholder or a combination of desires from various stakeholders. A higher score for a contextual preference indicates a higher level of importance of satisfying/performing the associated goal/task. This stands in comparison to the respective alternatives that either have contextual preferences with lower scores or those not mentioned at all. For instance in Figure 2, which shows a preference catalogue for our personal medication assistant, contextual preference $p_2$ and $p_4$ are respectively associated to $t_1$ and $g_3$, which are both alternatives to satisfying goal $g_1$. It becomes explicit that $g_3$ is preferred over $t_1$ when the patient location is *near_dispenser*, or even if both context instances in $p_2$ and $p_4$ hold. A similar notion is applied to the contextual preferences over softgoals. Softgoals do not always have equal importance. Different stakeholders in different contexts are interested in satisfying different subsets of softgoals, to which they also give different levels of importance. Referring to Figure 2, only three softgoals are given importance among all softgoals in the goal model. This is regardless of the type of patient illness,

```
p1 = (perform(t₁) • perform(t₅) • perform(t₇), patient_illness ∈
{dementia}, 9)
p2 = (perform(t₁), body_condition ∈ {tired, sick}, 3)
p3 = (perform(t₁), accompanying_people ∈ {alone} ∧
patient_activity ∈ {busy}, 4)
p4 = (satisfy(g₃), patient_location ∈ {near_dispenser}, 8)
p5 = (satisfy(g₃), accompanying_people ∈ {caregiver,
relatives}, 5)
p6 = (perform(t₈), weather ∈ {good}, 7)
p7 = (satisfy(sg₁), patient_illness ∈ {All}, 6)
p8 = (satisfy(sg₆), patient_illness ∈ {All}, 2)
p9 = (satisfy(sg₅), patient_illness ∈ {All}, 2)
```

**Figure 2: A preference catalogue.**

i.e., *patient_illness ∈ {All}*. The most important softgoal *sg₁* has the highest preference score of 6, both *sg₆* and *sg₅* with a preference score of 2, while all other softgoals are regarded as indifferent, i.e., their preference score is 0.

## 4 REQUIREMENTS VARIABILITY IN VARYING CONTEXTS AND PREFERENCES

We show how the candidate solutions in a goal model satisfy the given preference specification in various degrees. To simplify discussion in this section, we take a fragment of our personal medication assistant goal model as shown in Figure 3. That fragment is a subgraph that describes the goal *track medication* (*g₂*) which is AND-decomposed to: record intake (*g₅*), monitor medication effects (*g₆*), and inform relatives (*t₉*). Both *g₅* and *g₆* are further OR-decomposed to: *patient confirms intake* (*t₆*), *auto-confirm intake* (*t₅*), and *auto-monitoring of vital signs* (*t₇*), *manual monitoring* (*t₈*) respectively. Even in this small example, there can be several potential solutions for satisfying the root goal *g₂*. Each solution, which is composed of a set of leaf-level tasks, must satisfy the AND/OR structure of the root goal. For instance, the tasks [*t₆*, *t₅*] do not comprise a solution for the root goal because the monitoring of medication side effects is not achieved at all.

Given a preference catalogue *P*, we identify the contextual preference specifications from *P* that are relevant to a given situation. This situation is defined by a context instance which may refer to either case: a potential condition for a system-to-be or the current operational environment. In the former case, the context instance is explicitly expressed to explore the suitability of various behavioural designs. In the latter case, the context instance corresponds to the current context, that is, the context surrounding a running system. In such latter case, we assume a requirements model at runtime [25] that facilitates runtime adaptation, such as the goal-based service composition approach in [3]. The advances in sensor and context-aware technologies enable the capture of such context instances and various methods for capturing this (implicit) context are considerably discussed in the literature.

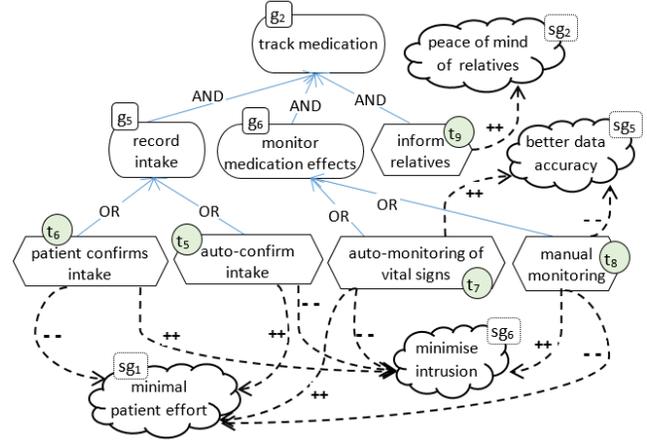

**Figure 3: A fragment of the personal medication assistant goal model.**

**Definition 6** (Imply relation between context instances). Given two context instances $cs_1 = (c_1^1,...,c_n^1)$ and $cs_2 = (c_1^2,...,c_n^2)$, $cs_2$ implies $cs_1$ if $\forall i, 1 \leq i \leq n, c_i^2 = c_i^1$ or $c_i^2 = All$.

A contextual preference *p* is relevant to a given situation when any context instance of *p* implies the context instance of the given situation. A context instance of *p* must be the same or more general than that of the given situation. Suppose a situation is described by the context assertion *con* = (*patient_activity ∈ {idle} ∧ patient_location ∈ {living_room} ∧ patient_illness ∈ {dementia} ∧ weather ∈ {good} ∧ body_condition ∈ {normal} ∧ accompanying_people ∈ {caregiver}*). Consequently, this situation is specified by the context instance *(idle, living_room, dementia, good, normal, caregiver)*. From the preference catalogue in Figure 2, we identify the relevant contextual preferences: *p1, p5, p6, p7, p8,* and *p9*. However, using the goal model in Figure 3, only *p1, p6, p7, p8,* and *p9* would be relevant, i.e., we exclude *p5* because it involves *g₃* which is not part of Figure 3. These relevant preferences are shown in Figure 4. The contextual preference *p1* is an explicit expression of prioritisation on the alternatives *t₅* and *t₇* for achieving goals *g₅* and *g₆* respectively. *p6* expresses the degree of preference to *t₈* for achieving *g₆*. *p7, p8*, and *p9* are the preferences for the preferred softgoals *sg₁, sg₆,* and *sg₅* respectively.

We now calculate the degree by which a solution satisfies the contextual preferences. We compute the *softgoal preference score* and *hardgoal preference score* of each solution. Then, we combine the scores to derive an overall *preference satisfaction degree* for each solution.

```
p1 = (perform(t₁) ∨ perform(t₅) ∨ perform(t₇), patient_illness
∈ {dementia}, 9)
p6 = (perform(t₈), weather ∈ {good}, 7)
p7 = (satisfy(sg₁), patient_illness ∈ {All}, 6)
p8 = (satisfy(sg₆), patient_illness ∈ {All}, 2)
p9 = (satisfy(sg₅), patient_illness ∈ {All}, 2)
```

**Figure 4: A set of applicable contextual preferences.**

## 4.1 Softgoal Preferences

To calculate the softgoal preference score, we regard the contribution links from a solution to the preferred softgoals. We apply the approach, set forth by the NFR framework [18] and further reused in [27] and [2], using quantitative estimations for assessing the positive or negative contribution of the candidate solutions to the softgoals. This approach considers every contribution link as evidence of the positive or negative satisfaction of a preferred softgoal.

**Definition 7** (Softgoal preference score). A softgoal preference score *sps(sol)* is the score derived for a solution *sol* with respect to satisfying the preferred softgoals.

*sps(sol) = contrib(sol, sg₁) + … + contrib(sol, sgₙ)*, where:
- *contrib(sol, sgᵢ) = percentPos(sol, sgᵢ) \* score(sgᵢ) - percentNeg(sol, sgᵢ) \* score(sgᵢ)*, is the contribution score of the solution *sol* to *sgᵢ*,
- each softgoal $sg_i$, $1 \leq i \leq n$, is a softgoal associated to a contextual preference *p*, and for any two softgoals $sg_i$, $sg_k$, $i \neq k$.

The function *percentPos(sol, sgᵢ)* refers to the percentage of the positive contributions ($\xrightarrow{++}$) with respect to the total number of contributions from *sol* to $sg_i$, while *percentNeg(sol, sgᵢ)* refers to the percentage of the negative contributions ($\xrightarrow{--}$) with respect to the total number of contributions from *sol* to $sg_i$. These functions are used to normalise the different number of contribution links upon softgoals. The function *score(sgᵢ)* is the priority degree defined for $sg_i$ in a contextual preference. Where $sg_i$ appears in multiple relevant contextual preferences, the highest defined score for $sg_i$ from among the relevant preferences is chosen for *score(sgᵢ)*.

We show in Table 2 the derived softgoal preference score of each candidate solution of our requirements problem illustrated in Figure 3. A matrix cell associated with a candidate solution *sol* and a softgoal *sg* captures the *contrib(sol, sg)*, which is the estimated preference score of *sol* with respect to *sg*. The last column of the matrix shows the total preference scores of each candidate solution, i.e., *sps(sol)*. For instance, *sps(a)* = 6 is the sum of the preference scores of solution *a*: [$t_5$, $t_7$, $t_9$] with respect to each preferred softgoal.

## 4.2 Hardgoal Preferences

We consider the relevant contextual preferences over some hardgoals in the goal model to derive the hardgoal preference score of each candidate solution.

**Definition 8** (Hardgoal preference score). A hardgoal preference score *hps(sol)* is the score derived for a solution *sol* with respect to the contextual preferences over the goal alternatives.

*hps(sol) = pref(sol, hg₁) + … + pref(sol, hgₙ)*, where:
- *pref(sol, hgᵢ) = score(hgᵢ)*, is the preference score of the solution *sol* to the alternative hardgoal $hg_i$,
- each $hg_i$, $1 \leq i \leq n$, is a hardgoal associated to a contextual preference *p*, and for any two hardgoals $hg_i$, $hg_k$, $i \neq k$.

The function *score(hgᵢ)* is the priority degree defined for $hg_i$ in a contextual preference. Where $hg_i$ appears in multiple relevant contextual preferences, the highest defined score for $hg_i$ from among the relevant preferences is chosen for *score(hgᵢ)*. For instance, contextual preference *p4* and *p5* in our preference catalogue in Figure 2 both define preferences for satisfying $g_3$. When both *p4* and *p5* become relevant in a given situation, *score(g₃)* = 8, which is the higher score of the two. Using the same goal model in Figure 3 and contextual preferences in Figure 4, we show in Table 3 the derived hardgoal preference score of each candidate solution. The middle column lists the *pref(sol, hg)* –the preference score of each preferred hardgoal *hg* that is satisfied in a particular solution *sol*. The last column totals the preference scores deriving the respective *hps(sol)*. For instance, *hps(a)* = 18 adds the preference scores associated with the preferred hardgoals, $t_5$: 9, $t_7$: 9, that are satisfied in the solution *a*: [$t_5$, $t_7$, $t_9$].

## 4.3 Preference Satisfaction Degree

Each solution satisfies the relevant contextual preference specifications to a different degree. By considering both the softgoal and hardgoal preference scores, we derive the preference satisfaction degree for each candidate solution. We add the softgoal and hardgoal preference scores *psd(sol) = sps(sol) + hpd(sol)*. Hence, given the context instance *(idle, living_room, dementia, good, normal, caregiver)* and the contextual preference specifications in Figure 2, the preference satisfaction degree of solution *a*, *b*, *c*, *d*, are 24, 14, 11, and 1 respectively.

In a different situation, such as for a patient without dementia: *(patient_activity ∈ {idle} ∧ patient_location ∈ {living_room} ∧ patient_illness ∈ {normal} ∧ weather ∈ {good} ∧ body_condition ∈ {normal} ∧ accompanying_people ∈ {caregiver})*, we define a context instance *(idle, living_room, normal, good, normal, caregiver)*. This would change the relevant contextual preferences to *p6*, *p7*, *p8*, and *p9*. Although preferences p7, p8, p9 to the respective softgoals $sg_1$, $sg_6$, $sg_5$ remain the same, only the hardgoal alternative $t_8$ is preferred, as defined in $p_6$. Consequently, we derive 6, 5, 2, 1 as the preference satisfaction degree of solution *a*, *b*, *c*, *d*, respectively. Should the context further change to *(idle, living_room, normal, bad, normal, caregiver)*, such as when the weather becomes bad, the satisfaction degree of the solution *a*, *b*, *c*, *d* would be 6, -2, 2, -6, respectively. The latter context instance would consider only the softgoal preferences because no hardgoal preference applies.

Table 2: Softgoal preference scores.

| Candidate solution (sol) | Preferred softgoals | | | sps(sol) |
|---|---|---|---|---|
| | $sg_1$ (score =6) | $sg_6$ (score =2) | $sg_5$ (score =2) | |
| *a*: [$t_5$, $t_7$, $t_9$] | 6 | -2 | 2 | 6 |
| *b*: [$t_5$, $t_8$, $t_9$] | 0 | 0 | -2 | -2 |
| *c*: [$t_6$, $t_7$, $t_9$] | 0 | 0 | 2 | 2 |
| *d*: [$t_6$, $t_8$, $t_9$] | -6 | 2 | -2 | -6 |

Table 3: Hardgoal preference scores.

| Candidate solution (*sol*) | Hardgoal preferences | hps(*sol*) |
|---|---|---|
| *a*: [$t_5$, $t_7$, $t_9$] | $t_5$: 9, $t_7$: 9 | 18 |
| *b*: [$t_5$, $t_8$, $t_9$] | $t_5$: 9, $t_8$: 7 | 16 |
| *c*: [$t_6$, $t_7$, $t_9$] | $t_7$: 9 | 9 |
| *d*: [$t_6$, $t_8$, $t_9$] | $t_8$: 7 | 7 |

As we have discussed in Section 2.2, different stakeholders have varying priorities across different situations. Our preference catalogue, for example as shown in Figure 2, captures both the different stakeholder priorities and the variations of such priorities. On the one hand, the catalogue can comprise individual preferences from various stakeholders, e.g., patient users, physicians, system default preferences, or organisational policies of the assistive medication service provider. On the other hand, expressing priorities as contextual preferences captures the variations of such priorities in different situations. For instance, the contextual preferences $p_1$, $p_2$, and $p_3$ express different prioritisations over the alternative task $t_1$. This can be interpreted as either the preferences of different stakeholders to $t_1$, or the varying preference to $t_1$ that depends on context. The latter can consider the preferences as the changing levels of interest of one stakeholder for $t_1$.

Moreover, a contextual preference specification in the preference catalogue can be added, removed, or updated whenever necessary. Suppose the stakeholders decide that among the three preferred softgoals, maintaining privacy should become the most important as long as the patient has no dementia nor MCI. Otherwise, the originally set preferences remain. We can resort to either two options. The first option is keeping the present softgoal preference specifications unchanged and adding a new one that gives the softgoal *minimise intrusion* ($sg_6$) with a score higher than the rest. For example, we add $p_{10}$ = *(satisfy(sg_6), patient_illness ∈ {normal}, 8)*. Hence, in a situation that defines *patient_illness = normal*, the contextual preference $p_8$ and $p_{10}$ would become relevant, i.e., both apply to the situation and softgoal $sg_6$. In this case, $score(sg_6) = 8$, that is, being the higher score between the two. The second option adds two new contextual preferences. For example, we add $p_{10}$ = *(satisfy(sg_6), patient_illness ∈ {normal}, 6)* and $p_{11}$ = *(satisfy(sg_1), patient_illness ∈ {normal}, 2)*. This is consistent with the scoring of the existing softgoal preference specifications, i.e., similarly assigning 6 as score for the most important one. Likewise, we need to update the context assertion of $p_7$ and $p_8$, i.e., $p_7$ = *(satisfy(sg_1), patient_illness ∈ {dementia, MCI}, 6)* and $p_8$ = *(satisfy(sg_6), patient_illness ∈ {dementia, MCI}, 2)*. The context assertion is now altered from *patient_illness ∈ {All}* to *patient_illness ∈ {dementia, MCI}* for both contextual preferences. Therefore, in both options, when the patient is normal, i.e., *patient_illness = normal*, the candidate solutions that may get higher softgoal preference scores are those that positively contribute to the softgoal *minimise intrusion* ($sg_6$). Considering the second option and the context instance *(idle, living_room, normal, good, normal, caregiver)* which reflects *patient_illness = normal*, we derive for our goal model in Figure 3 preference satisfaction degree -2, 5, 2, 9 for solutions *a*, *b*, *c*, and *d*, respectively.

## 5 REASONING ABOUT CONTEXTUAL PREFERENCES

We exploit the powerful method for declarative knowledge representation and reasoning provided by Answer Set Programming (ASP) [14] to support automation of the techniques described in the previous section. We develop a prototype reasoning tool that extends DLV [2] –a state-of-the-art ASP implementation. We translate both the goal model formalisms and contextual preference specifications into a disjunctive logic program appropriate for a DLV input file. Overall, our tool takes as input the goal model, the contextual preference specifications, and the context instance describing a situation. Then, it returns the derived alternative solutions to the goal model problem. The solutions are ranked by preference satisfaction degree.

We refer to the goal model in Figure 1 and apply the preference catalogue in Figure 2. Assuming a context instance (*busy, living_room, dementia, good, tired, caregiver*) that defines respective values for the context elements (*patient_activity, patient_location, patient_illness, weather, body_condition, accompanying_people*), the resulting ranked solutions are shown in Table 4. In this particular situation, the relevant contextual preferences are $p_1$, $p_2$, $p_5$, $p_6$, $p_7$, $p_8$, and $p_9$. Solutions that include most of the highly preferred hardgoals and those that positively contribute to most of the preferred softgoals get better scores. We look at the solution with the highest satisfaction degree: [$t_1$, $t_5$, $t_7$, $t_9$], that is, the *optimal solution*. It contains $t_1$, $t_5$, and $t_7$, which are the highly preferred alternatives when the patient has dementia, as specified by $p_1$. The contextual preference $p_2$ also applies, but its effect is overshadowed by $p_1$, since both preferences associate with $t_1$ (see Section 4.2).

Regarding the softgoal preferences, the optimal solution has two (non-negated) negative contributions to the preferred softgoal $sg_6$, but the (non-negated) positive contributions to the preferred softgoals $sg_1$ and $sg_5$ have more weight, thus, still obtaining a significant positive softgoal preference score.

**Table 4: Ranked solutions.**

|   | Candidate solution | Preference satisfaction degree |
|---|---|---|
| 1 | [$t_1$, $t_5$, $t_7$, $t_9$] | 330 |
| 2 | [$t_1$, $t_6$, $t_7$, $t_9$] | 260 |
| 3 | [$t_1$, $t_5$, $t_8$, $t_9$] | 250 |
| 4 | [$t_2$, $t_3$, $t_4$, $t_5$, $t_7$, $t_9$] | 230 |
| 5 | [$t_1$, $t_6$, $t_8$, $t_9$] | 180 |
| 6 | [$t_2$, $t_3$, $t_4$, $t_6$, $t_7$, $t_9$] | 160 |
| 7 | [$t_2$, $t_3$, $t_4$, $t_5$, $t_8$, $t_9$] | 150 |
| 8 | [$t_2$, $t_3$, $t_4$, $t_6$, $t_8$, $t_9$] | 80 |

---

[2] http://www.dlvsystem.com/

Table 5: Results of running our tool for five different goal models.

| | $N_{HG}$ | $N_{SG}$ | $N_{CL}$ | $N_{VP}$ | $N_{CP}$ | $N_{Sol}$ | $T_{NS}$ | $T_{PR}$ | $T_{OS}$ | $T_{FNS}$ | $T_{FOS}$ |
|---|---|---|---|---|---|---|---|---|---|---|---|
| 1 | 15 | 6 | 13 | 3 | 13 | 8 | 0.3 | 0.0 | 0.4 | 0.3 | 0.4 |
| 2 | 31 | 12 | 26 | 6 | 26 | 64 | 0.5 | 0.1 | 0.6 | 0.4 | 0.6 |
| 3 | 46 | 18 | 39 | 9 | 39 | 512 | 1.2 | 0.7 | 2.0 | 0.4 | 1.9 |
| 4 | 61 | 24 | 52 | 12 | 52 | 4096 | 7.0 | 7.9 | 14.9 | 0.4 | 13.6 |
| 5 | 76 | 30 | 65 | 15 | 65 | 32768 | 68.2 | 104.2 | 172.4 | 0.4 | 160.7 |

**Legend** (all times are in seconds)
$N_{HG}$: number of hardgoals, $N_{SG}$: number of softgoals, $N_{CL}$: number of contribution links, $N_{VP}$: number of variability points, $N_{CP}$: number of contextual preferences, $N_{Sol}$: number of solutions, $T_{NS}$: time to derive all non-optimised solutions, $T_{PR}$: time for contextual preference reasoning, $T_{OS}$: time to derive all optimised solutions, $T_{FNS}$: time to derive the first solution (may be a sub-optimal one), $T_{FOS}$: time to derive the optimal solution

A solution's positive (resp. negative) contribution to a softgoal is significant only if there is no negative (resp. positive) contribution from the same solution that negates it. For instance, the solution [$t_2$, $t_3$, $t_4$, $t_6$, $t_7$, $t_9$] has one positive contribution to $sg_1$ (from $t_7$), but this is negated by the negative contribution from goal $g_3$ upon performing the tasks $t_2$, $t_3$, and $t_4$. Moreover, the resulting ranking changes with the change in context. We expect solutions that contain $t_2$, $t_3$, and $t_4$ to get the better scores when a patient without dementia is near the medicine dispenser as reflected in the context instance (*busy, near_dispenser, normal, good, tired, caregiver*).

We applied our approach on five goal models with increasing sizes. Table 5 summarises the results of running our tool over these models on a machine with 3.19 Ghz CPU and 16 GB RAM. Row 1 shows results for our original goal model in Figure 1. We cloned our original goal model to itself, to obtain those goal models with larger sizes, a similar approach done in [28]. Simultaneously, we cloned the contextual preferences in Figure 2 to generate preferences that apply to each goal model. The times in Table 5 show the mean of 20 runs of our tool over the same goal model and set of contextual preferences. The times are rounded-off to the nearest tens. All times are directly observed from running our tool except the time for contextual preference reasoning which we derived as $T_{PR} = T_{OS} - T_{NS}$. Our results show an exponentially increasing computation time to derive an optimal solution, as the goal model problem grows in size. Overall, although comparisons with other approaches are yet to be done, our tool performs considerably well in the test cases having up to 4,096 alternative solutions. We believe that most realistic small to medium-sized goal model problems would fall within or closely above such potential number of solutions. However, this performance is still less sufficient particularly for large-sized goal model problems, since it took 160.7 seconds to find the optimal solution among 32,768 candidate solutions.

## 6 RELATED WORK

The importance of capturing preferences in requirements modelling has been extensively recognised in the literature. With the potentially large space of candidate solutions to the requirements problem, stakeholder preferences have been used as criteria for comparison. There are two main types of preference specification: the qualitative specification, such as in [22], [11], [21], and [7]; and the quantitative specification, as like in [24], [16], and [20]. We first look at the qualitative ones. In the novel requirements modelling framework Techne [11], requirements and their relations are represented in a graph called r-nets. A preference is represented as a node in the model to express a binary relation between two requirements. An arrow line is drawn from the more preferred requirement to the preference node, and from the preference node to the less preferred requirement. Another qualitative preference specification framework is the CI-net [21], which captures complex binary preferences tradeoffs between multiple optional goals and ranks those goals based on the preferences. Both frameworks aim to find a set of highly preferred optional goals to be set as criteria in comparing candidate solutions. Moreover, the pQGM framework [7] extends the qualitative goal modeling framework in [22] to accommodate priorities expressed among i) optional goals and ii) goal alternatives. This framework utilises r-nets to express a binary preference relationship between goal model elements.

Alternatively, the quantitative preference specifications are more elaborate by using numerical weights. The work in [24] quantitatively measures the impact of alternative solutions on the degree of satisfaction of softgoals. Evaluating such impacts is used in the selection of a most preferred alternative. Liaskos et al. [16] proposed a goal modelling framework that considers numerical preferences expressed over optional goals. A planner is utilised to identify solutions that would best satisfy the specified preferences. Nguyen et al. [20] proposed the CGM-Tool, a realisation of an expressive modelling language that provides constructs to express both qualitative and quantitative preferences. In addition to the binary preference relations between goal model elements, the CGM-Tool can also express numerical preferences as i) attributes (i.e., penalties or rewards) for goals and ii) numerical objectives to optimise. Our approach similarly captures stakeholders' preferences to reason about alternative solutions. However, we believe that preferences are not fixed and may vary according to context. Hence, we focus on representing and reasoning about contextual preferences, which has been given less attention in the RE literature. None of the aforementioned works have explicitly captured the varying preferences attributed to contextual changes.

## 7 CONCLUSION AND FUTURE WORK

We introduced a goal-based requirements variability framework for modelling and reasoning about contextual preferences. The framework presents a contextual preference specification that extends the traditional requirements preferences with contextual information. Context is modelled as a set of environment elements. Each of the elements takes a value from its corresponding domain to define a context instance. The context instance describes the situation expressed by the context assertion that annotates a requirements preference. Hence, a contextual preference specification defines the situation when a particular level of importance, which is assigned as a numerical score, is given to an alternative hardgoal or a softgoal. A goal model with the contextual preference specification is further translated into a disjunctive logic program. The state-of-the-art DLV, which is a powerful implementation for knowledge representation and reasoning, is utilised to derive alternative solutions to the requirements problem. The derived solutions are further ranked according to their degree of satisfying the contextual preferences.

Our framework is useful in exploring potential system designs to be operationalised, that is, considering the satisfaction of the varying fitness criteria posed by stakeholders. While our framework aims to find one most optimal solution, it can also provide multiple potential solutions, i.e., the top-ranked (optimal) solutions. Providing multiple solution alternatives might give flexibility to the systems analysts in making their decisions. For our on-going and future work, we are looking for case studies to integrate our approach with industrial requirements analysis practices of which we expect goal model problems with more complex contextual preferences. We also plan to optimise the associated reasoning approach for our tool to scale well with large-sized goal models and complex contextual preferences. In addition, we plan to address the following problems: dealing with conflicts between contextual preferences, creating and combining contextual preference sets that would allow preferences to be grouped, e.g., according to stakeholders/users, and enabling (visual) traceability of the contextual preferences and the resulting goal model solutions. Furthermore, our quest for a comprehensive context-aware requirements variability framework drives us to continue exploring other relationships between context and requirements variability.

## REFERENCES

[1] Agrawal, R. and Wimmers, E.L., 2000. A Framework for Expressing and Combining Preferences. In *Proceedings of the 2000 ACM SIGMOD International Conference on Management of Data* (Dallas, Texas, USA, 2000), ACM, 335423, 297-306.
[2] Ali, R., Dalpiaz, F., and Giorgini, P., 2013. Reasoning with Contextual Requirements: Detecting Inconsistency and Conflicts. *Information and Software Technology 55*, 1, 35-57.
[3] Cavallaro, L., Sawyer, P., Sykes, D., Bencomo, N., Val, #233, and Issarny, R., 2012. Satisfying Requirements for Pervasive Service Compositions. In *Proceedings of the 7th Workshop on Models@run.time* (Innsbruck, Austria, 2012), ACM, 2422522, 17-22.
[4] Dalpiaz, F., Franch, X., and Horkoff, J., 2016. iStar 2.0 Language Guide. *arXiv preprint arXiv:1605.07767*.
[5] Dey, S. and Lee, S.-W., 2017. REASSURE: Requirements Elicitation for Adaptive Socio-technical Systems using Repertory Grid. *Information and Software Technology 87*(2017/07/01/), 160-179.
[6] Dourish, P., 2004. What we talk about when we talk about context. *Personal Ubiquitous Comput. 8*, 1, 19-30.
[7] Ernst, N.A., Mylopoulos, J., Borgida, A., and Jureta, I., 2010. Reasoning with Optional and Preferred Requirements. In *Conceptual Modeling-ER 2010* Springer, 118-131.
[8] Finkelstein, A. and Savigni, A., 2001. A Framework for Requirements Engineering for Context-aware Services. In *Proceedings of the 1st International Workshop From Software Requirements to Architectures (STRAW '01)* (Ontario, Canada, 2001), IEEE Computer Society Press.
[9] Galster, M., Weyns, D., Tofan, D., Michalik, B., and Avgeriou, P., 2014. Variability in software systems—a systematic literature review. *Software Engineering, IEEE Transactions on 40*, 3, 282-306.
[10] Henricksen, K., Indulska, J., and Rakotonirainy, A., 2006. Using context and preferences to implement self-adapting pervasive computing applications. *Software: Practice and Experience 36*, 11-12, 1307-1330.
[11] Jureta, I.J., Borgida, A., Ernst, N.A., and Mylopoulos, J., 2010. Techne: Towards a New Generation of Requirements Modeling Languages with Goals, Preferences, and Inconsistency Handling. In *18th IEEE International Requirements Engineering Conference*, 115-124.
[12] Kostavelis, I, Giakoumis, D., Malasiotis, S., and Tzovaras, D., 2016. RAMCIP: Towards a Robotic Assistant to Support Elderly with Mild Cognitive Impairments at Home. In *Pervasive Computing Paradigms for Mental Health: 5th International Conference, MindCare 2015, Milan, Italy, September 24-25, 2015, Revised Selected Papers*, S. SERINO, A. MATIC, D. GIAKOUMIS, G. LOPEZ and P. CIPRESSO Eds. Springer International Publishing, Cham, 186-195.
[13] Lamsweerde, A.V., 2000. Requirements engineering in the year 00: a research perspective. In *Proceedings of the 22nd International Conference on Software Engineering* (Limerick, Ireland, 2000), ACM, 337184, 5-19.
[14] Leone, N. and Ricca, F., 2015. Answer Set Programming: A Tour from the Basics to Advanced Development Tools and Industrial Applications. In *Reasoning Web. Web Logic Rules: 11th International Summer School 2015, Berlin, Germany, July 31- August 4, 2015, Tutorial Lectures.*, W. FABER and A. PASCHKE Eds. Springer International Publishing, Cham, 308-326.
[15] Liaskos, S., Lapouchnian, A., Yu, Y., Yu, E., and Mylopoulos, J., 2006. On Goal-based Variability Acquisition and Analysis. In *14th IEEE International Requirements Engineering Conference (RE'06)*, 79-88.
[16] Liaskos, S., Mcilraith, S.A., Sohrabi, S., and Mylopoulos, J., 2011. Representing and reasoning about preferences in requirements engineering. *Requirements Engineering 16*, 3, 227.
[17] Metzger, A. and Pohl, K., 2014. Software product line engineering and variability management: achievements and challenges. In *Proceedings of the on Future of Software Engineering* ACM, 70-84.
[18] Mylopoulos, J., Chung, L., and Nixon, B., 1992. Representing and using nonfunctional requirements: A process-oriented approach. *IEEE Transactions on Software Engineering 18*, 6, 483-497.
[19] Negri, P.P., Souza, V.E.S., De Castro Leal, A.L., De Almeida Falbo, R., and Guizzardi, G., 2017. Towards an Ontology of Goal-Oriented Requirements. In *Proceedings of the 20th Ibero-American Conference on Software Engineering* (Buenos Aires, Argentina, 2017).
[20] Nguyen, C.M., Sebastiani, R., Giorgini, P., and Mylopoulos, J., 2016. Multi-objective reasoning with constrained goal models. *Requirements Engineering*, 1-37.
[21] Oster, Z.J., Santhanam, G.R., and Basu, S., 2015. Scalable modeling and analysis of requirements preferences: A qualitative approach using CI-Nets. In *2015 IEEE 23rd International Requirements Engineering Conference (RE)*, 214-219.
[22] Sebastiani, R., Giorgini, P., and Mylopoulos, J., 2004. Simple and minimum-cost satisfiability for goal models. In *International Conference on Advanced Information Systems Engineering* Springer, 20-35.
[23] Stefanidis, K., Pitoura, E., and Vassiliadis, P., 2011. Managing contextual preferences. *Information Systems 36*, 8, 1158-1180.
[24] Subramanian, C.M., Krishna, A., and Kaur, A., 2016. Sensitivity Analysis of the i* Optimisation Model. *Journal of Software 11*, 1, 10-27.
[25] Szvetits, M. and Zdun, U., 2016. Systematic literature review of the objectives, techniques, kinds, and architectures of models at runtime. *Software & Systems Modeling 15*, 1, 31-69.
[26] Van Lamsweerde, A., 2001. Goal-oriented requirements engineering: A guided tour. In *Fifth IEEE International Symposium on Requirements Engineering* IEEE, 249-262.
[27] Van Lamsweerde, A., 2009. Reasoning About Alternative Requirements Options. In *Conceptual Modeling: Foundations and Applications: Essays in Honor of John Mylopoulos*, A.T. BORGIDA, V.K. CHAUDHRI, P. GIORGINI and E.S. YU Eds. Springer Berlin Heidelberg, Berlin, Heidelberg, 380-397.
[28] Wang, Y., Mcilraith, S.A., Yu, Y., and Mylopoulos, J., 2007. An automated approach to monitoring and diagnosing requirements. In *Proceedings of the 22nd IEEE/ACM international Conference on Automated Software Engineering* (Atlanta, Georgia, USA, 2007), ACM, 1321675, 293-302.
[29] Yu, E.S.K., 1997. Towards modelling and reasoning support for early-phase requirements engineering. In *Third IEEE International Symposium on Requirements Engineering*, 226-235.